# Evolution of mass, surface layer composition and light scattering of evaporating, single microdroplets of SDS/DEG suspension. Shrinking droplet surface as the micelles generator.

Maciej Kolwas, Daniel Jakubczyk[*], Justice Archer[1], Tho Do Duc[2]

Institute of Physics, Polish Academy of Sciences

al. Lotników 32/46, PL-02668 Warsaw, Poland

**Abstract**

We study the evolution of the mass of evaporating single microdroplets of sodium dodecyl sulphate (SDS) / diethylene glycol (DEG) mixture. First, we recognise and deconvolute the influence of residual water evaporation [Kolwas *et al*. Soft Matter 2019;15:1825], which accelerates the composite droplet evaporation, a simple exponential decay of the evaporating droplet surface change rate. This enables us to study the influence of SDS concentration on the composite droplet evaporation. Next, we establish a simple relationship between the average SDS concentration and the droplet evaporation rate to enable the study of the evolution of SDS concentration at the droplet surface. The oscillatory nature of surface SDS concentration indicates cyclic changes in the surface monolayer associated with the cyclic creation of vesicles (micelles) at the surface. The model we developed, allows determination of SDS critical micelles concentration (CMC) in DEG as 60±2 mM.

## 1. Introduction

The physical, surface-thermodynamic properties of surfactants have been the subject of research for a long time since surfactants have been used in a variety of applications – from everyday to purely scientific, e.g. in biochemistry, biotechnology and medicine (see e.g. [1], [2], [3], [4], [5], [6], etc.). Sodium dodecyl sulphate (SDS), which we used in this study, is a particularly widely used representative.

Specific properties of surfactants are associated with their ability to form a surface layer consisting of monomers – Langmuir film. The monomer concentration increases with the (average) concentration of surfactant in the volume of the mixture, up to the monolayer saturation [7] at so called Critical Micelles Concentration (CMC). Further increase of the average surfactant concentration leads to the increase of micelles concentration (e.g. [8], [9], [10]). Therefore, for concentrations higher than CMC, an SDS-containing mixture resembles a colloidal suspension rather than solution. The surfactant (SDS) concentration in the surface monolayer stays saturated and constant, if only the surface area stays constant – the surface is not compressed.

As liquids in nature and in technology are very often encountered in the form of droplets they should be investigated in such form as well. With the decrease of droplet size, the ratio be-tween its surface area and its volume grows. It makes surface phenomena essential for small droplets – the thermodynamics of micro- and nanodroplets is strongly influenced by surface phenomena [11].

[*] Corresponding author. E-mail address: Daniel.Jakubczyk@ifpan.edu.pl
[1] Present address: Bristol Aerosol Research Centre, School of Chemistry, Cantock's Close, University of Bristol, Bristol, BS8 1TS, UK. E-mail: archerjk@yahoo.com
[2] Present address: School of Engineering Physics, Hanoi University of Science and Technology, Ha Noi, Vietnam. E-mail: tho.doduc@hust.edu.vn

During evaporation of solvent/dispersion medium from a droplet containing a surfactant, the surfactant surface monolayer is compressed due to shrinking of surface area. At low surfactant concentration it behaves just as the Langmuir monolayer [5]. Above the CMC, the saturated monolayer must fold/buckle or break, undergoing the transition from a 2D to a 3D structure [12], [13]. The state of the surface layer [14] should influence the evaporation. Therefore, in principle, it should be possible to investigate the state of the surface layer by following the evaporation rate of the droplet and we report such investigations in this paper. Our method augments in this respect other methods reported in literature. Usually it is only possible to follow the changes of the average concentration of components in a composite droplet, either by measuring the evaporation rate [15], [16] or by measuring the changes of the refractive index [17], [18]. Rainbow refractometry [19] can also provide information on the average refractive index profile in the droplet and from that infer the average composition profile, however, it is extremely sensitive to any droplet non-sphericity [20]. Brewster Angle Microscopy (BAM) [14] – a technique eagerly used for investigation of surface-active compounds – can hardly be applied to levitating microdroplets, while Small Angle Neutron Scattering (SANS) has been applied only to microfluidic droplets so far [21].

We present an experimental study of the evolution of the surface and its structure of a free microdroplet of SDS/diethylene glycol/water solution, driven by the evaporation of solvents. We investigated single droplets with well-defined initial concentration of SDS mixture with DEG, levitating in an electrodynamic trap.

The surface area evolution was obtained from measurement of the droplet mass with electrostatic weighting, calibrated with the Mie Scattering Look-up Table Method (MSLTM, see [22] for details). We also analysed optical morphology/structure dependent resonances (MDRs) manifesting in the intensity of the scattered light, which we observed around the right angle. It allowed us to examine the evolution of the microdroplet surface smoothness/structure.

## 2. Experimental details

We prepared the solutions of SDS in diethylene glycol (DEG) in ambient air at room temperature. Prior to sample preparation, DEG was kept under vacuum. Since glycols are highly hygroscopic, we estimate that after preparation there was ≤4 wt% of water in the solution due to the ambient atmosphere moisture and contact with plastic syringes (plastics can contain up to several wt% of water [23]). Vacuum distillation of the solution under standard temperature was helping a little with this issue. However, we avoided storing and processing of the SDS solution for long, since it has a known tendency to age and undergo internal (phase) changes [24]. Having all this in mind, in the presented experiments we did not strive to make as dry SDS solutions as possible, but rather accounted for the contamination with water in post-processing (see Section 3.1).

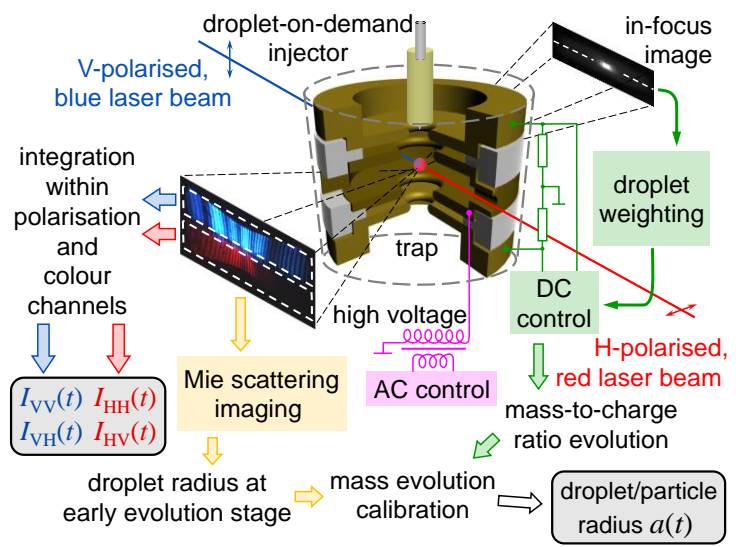

Fig. 1. Schematic diagram of the experimental setup, experiment control and data acquisition.

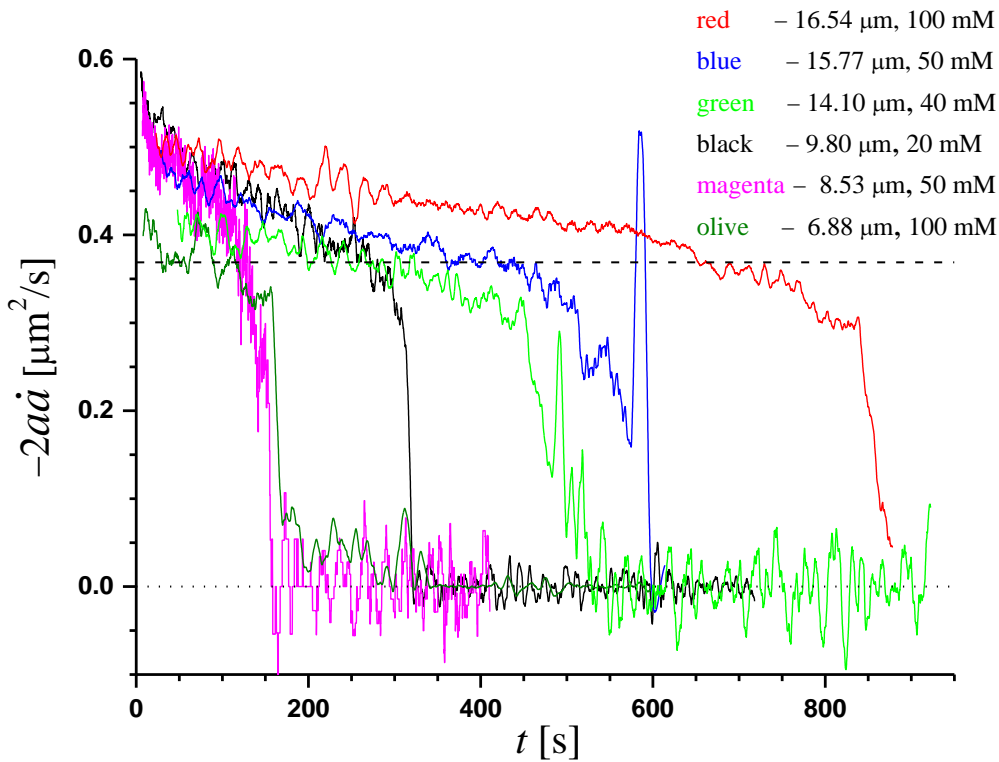

Fig. 2. Evaporation rate $-da^2/dt$ (expressed as $-2a\dot{a}$) of the investigated droplets obtained from the measured droplet mass evolution. The pure DEG evaporation rate is indicated as the dashed black line. Initial values of the SDS concentration and the initial radii are shown in the legend. The color coding of the curves/experimental runs introduced here is kept throughout the text.

Single, charged microdroplets of solution were introduced into the 3D electrodynamic trap with a droplet-on-demand injector (see Fig. 1 and [25]). The initial radius of a microdroplet was controllable to a certain extent with the injection driving pulse parameters (amplitude, shape, length and timing). Electrodynamic trapping is an established technique of levitating a charged particle/droplet with a combination of AC and DC electric fields [22]. The vertical (along the axis of the trap) position of a droplet was stabilized at the trap centre with the DC voltage dependent on the observed position of the droplet. Following the changes of this voltage enabled measuring the evolution of the droplet mass-to-charge ratio and finally finding the evolution of the microdroplet radius $a(t)$. The accuracy of electrostatic weighting is limited mainly by the discretization of the droplet position reading (with the CCD sensor) and is estimated at ~1 %. Finding the evaporation rate $-2a\dot{a}(t)$ requires however calculating the $\dot{a}$ derivative numerically. This introduces significant noise with central frequency associated with the camera frame rate – 30 fps in the presented case. Since the studied phenomena are much slower, it is legitimate to smooth out the high frequency noise. Since we did not observe any losses of the droplet charge, the radius was simply calibrated at the beginning of the evo-lution with MSLTM – radius of a nearly homogeneous droplet was independently measured by analysing the scattered light intensity pattern with the help of Mie theory. Since SDS does not evaporate, it is possible, with the binary liquid evaporation model at hand [16] to find an exact radius value. The task was facilitated by the observation that the density of the solution can be assumed constant, as far as the density of DEG is 1.12 g/cm$^3$, of SDS is 1.01 g/cm$^3$, and of water is 1 g/cm$^3$.

The MDRs were observed in the intensity of scattered light integrated over the field of view (FoV; 16$^o$ angle around the right angle in the scattering plain, ±7$^o$ elevation). Such integration increased the signal to noise ratio and enabled analysing the fine details of scattering resonant features.

We studied microdroplets of SDS/DEG solutions with 4 initial SDS concentrations $c_{SDS}$ = 20, 40, 50 and 100 mM. The initial droplet radius $a_0$ ranged from ~6 μm to ~17 μm, as exemplified in Fig 2.

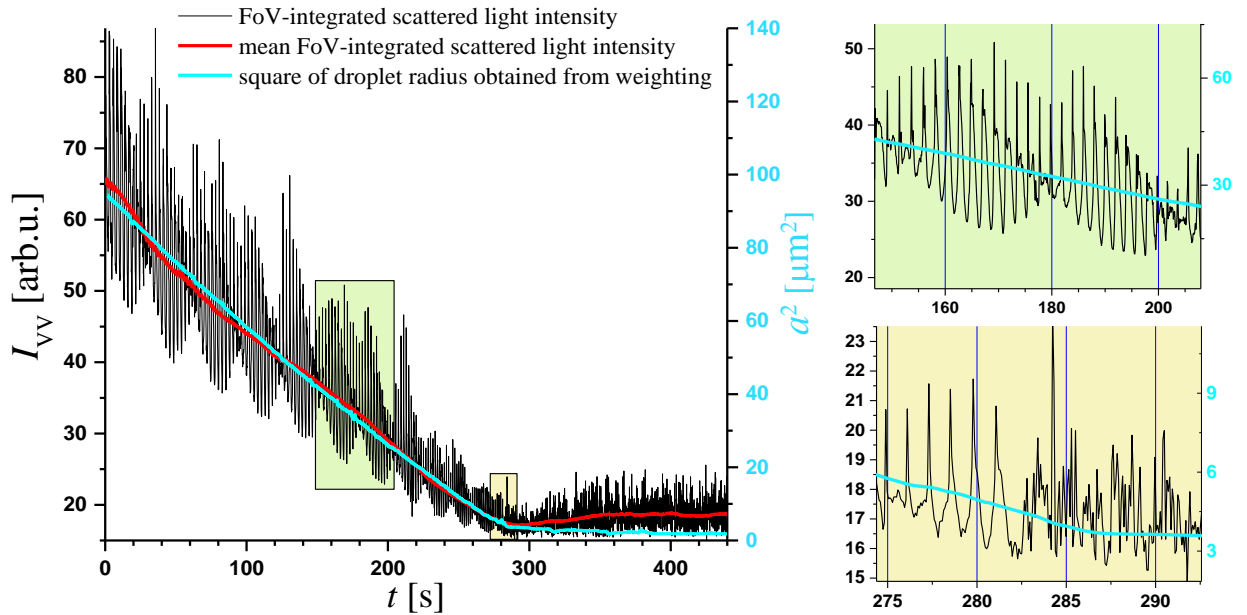

Fig. 3. Evolution of the FoV-integrated intensity of the scattered light for a droplet with $a_0$ = 9.8 μm and $c_{SDS}$ = 20 mM (black line). A distinct regular pattern associated with the MDRs of a homogeneous droplet (see magnification in the upper right panel) transforms at $t \cong 282$ s into an irregular sequence of spikes (magnified in the lower right panel). Cyan line represents the evolution of $a^2$ evaluated from the droplet weighting. Red line represents the mean intensity of the scattered light and can be perceived as a measure of the optical cross-section or square of "optical radius".

## 3. Discussion of the experimental data

Seven examples of microdroplet evaporation dynamics are shown in Fig. 2 in the form of temporal evolution of the radius-square change rate, proportional to the rate of change of the droplet surface $-2a\dot{a}(t)$. The color coding of the curves/experimental runs introduced here is kept throughout the text. Three distinct regions of evolution can be identified:

(i)  a slow decrease of evaporation rate – due to the evaporation of residual water;
(ii) a much steeper decrease followed by;

(iii) a nearly constant evaporation rate near zero, when the remnants of DEG evaporate

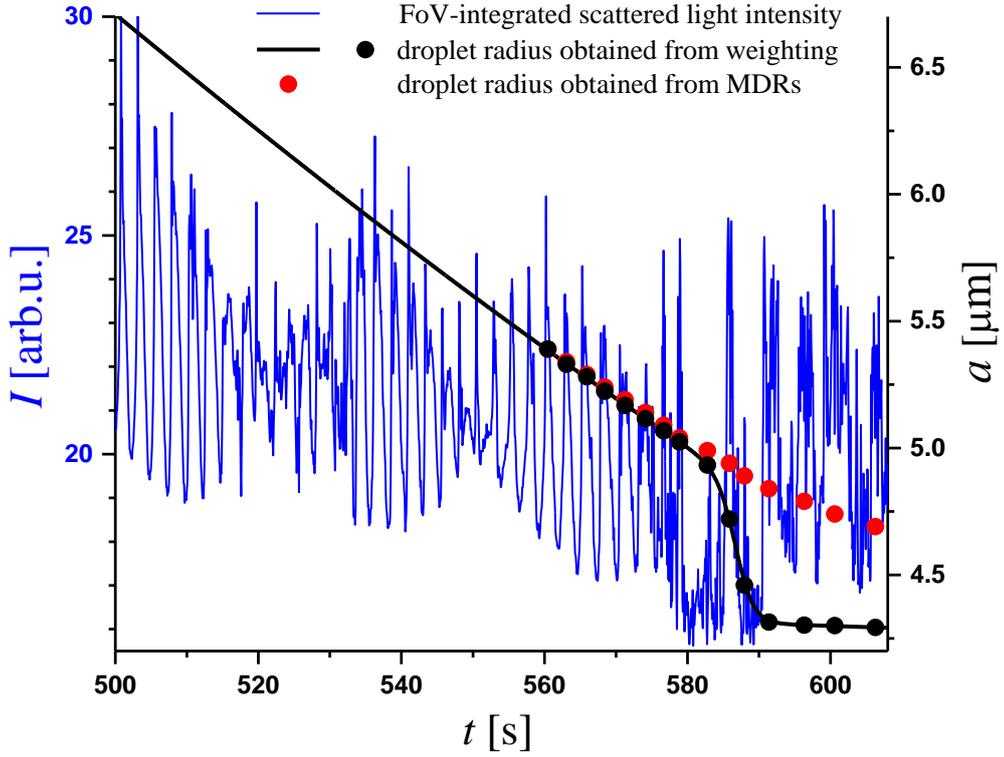

Fig. 4. The final part of the temporal evolution of the FoV-integrated scattered light intensity (blue solid line) and radius (black solid line, red and black circles) of a droplet with the initial SDS concentration of 50 mM and the initial radius of 15.77 μm. The MDR structure is devastated at ~515 s, and revived at ~535 s. After ~575 s MDRs again become irregular and finally disappear. The "optical radius", measured with the intervals between the MDR maxima is represented with red circles and radius obtained with electrostatic weighting is represented with black solid line with black circles.

and the droplets transform into solid SDS micro-objects.
Additionally, for the droplet evolutions labelled with the light green and blue lines, an "explosion" of evaporation rate occurs at ~500 s and ~600 s respectively. This suggests fracturing of a non-elastic surface layer and spillage of the liquid from below onto the rough surface, which dramatically increases the evaporation rate. For the other droplets, the evaporation rate evolutions are only slightly modulated. This modulation is the subject of our investigation, as we know from our earlier studies (e.g. [26]) that it may convey information about the surface structure of the composite droplets.

Generally, the FoV-integrated scattered light intensity temporal patterns correspond to two main states of droplet surface. The first pattern is associated with scattering on a droplet with smooth surface and corresponds to a manifold of MDRs of a spherical resonator – homogeneous droplet (e.g. [27], [28]). At a certain point a rapid pattern transformation occurs. The second pattern can be associated with scattering on a rough (solid) surface, which has a devastating influence upon the mode structure. An example evolution of the FoV-integrated intensity of the scattered light on the droplet ($a_0 = 9.8$ μm, $c_{SDS} = 20$ mM) is shown in Fig. 3. Cyan line, representing the evolution of $a^2$ evaluated from the droplet weighting (mass evolution), illustrates that the mean intensity of the scattered light is a nearly linear function of the surface (or cross-section) area (compare the "radius-square law" [29]). It is worth noticing (in the lower right panel of Fig. 3) that the moment when the evaporation slows down is retarded by a few seconds in respect to the moment when the irregularities in the scattering intensity begin – though the surface roughness develops, the evaporation is not impeded at first. Only a few seconds later it slows down – SDS is sealing the surface. The mean intensity of the scattered light (red line), which can be perceived as a measure of the optical cross-section (square of "optical radius"), is slightly increasing after the evaporation slow-down in contrast to slowly decreasing $a^2$ evaluated from the droplet weighting. This seems to indicate the increase of the droplet surface area due to the interface corrugation.

An exceptional behaviour – a revival of MDRs – was spotted for a bigger droplet with $a_0$ = 15.77 μm and initial SDS concentration $c_{SDS}$ = 50 mM (Figs. 2 and 4). The devastation of the MDRs at $t \cong 515$ s is for some reason followed at $t \cong 535$ s by their revival. It might be expected that the developing surface roughness is reversed – the droplet surface is re-smoothing – until $t \cong 575$ s. Then the final drying begins and MDRs are destroyed altogether. The re-smoothing at $t \cong 535$ s seems not to be connected with bursting of surface shell of SDS, such as at $t \cong 600$ s, but rather with the shell reconfiguration towards higher symmetry. It can also be noticed that the "optical radius", measured with intervals between the MDR maxima (red squares in Fig. 4), decreases slower than the radius obtained from the droplet weighting (black line with black circles). This seems to suggest the creation of large SDS structures ("roughness") on the droplet surface.

### *3.1. Influence of water on drying droplet of colloidal suspension.*

The evaporation of SDS/DEG solution/colloidal suspension is controlled mainly by growing concentration of SDS. However, at the beginning, it may be also influenced by the presence/evaporation of contaminating water, since DEG is highly hygroscopic. In order to study just the evaporation of SDS/DEG solution/suspension, we should recognize and eliminate the influence of water first. The observed phenomenon may be viewed as distillation at microscale – though both components evaporate simultaneously and mutually influence their evaporation, the more volatile component (water) leaves first and we just see the end of this process. Since the water/DEG mixture is not azeotropic and the nitrogen atmosphere was nearly perfectly dry, the final content of water in the droplet was negligible.

In our previous work [16] we studied evaporation of free droplets of binary liquid mixtures (hygroscopic liquids in particular) under diverse conditions and we have found exact analytical solutions for the evolution of droplet evaporation rate. Using these solutions in full is rather inconvenient, but the implications of these should enable us to easily circumvent the effects of contamination with water. The cases discussed in this work can be qualified as evaporation of water-contaminated liquid in dry atmosphere. Such evaporation is (initially) faster than for the pure liquid (DEG in this case, as can be seen in Fig. 2). At this stage of evolution the surface change rate (evaporation rate) follows a single-exponential decay (compare Eqns. 26 and 27 in [16]) converging to the evaporation rate of pure DEG:

$$\frac{d}{dt}S = 8\pi a \dot{a} = -A\exp\left(-\frac{t}{\tau}\right) + \left(\frac{d}{dt}S\right)_{DEG} \tag{1}$$

where $A$ and $\tau$ (water content dwell time) are constant for each droplet. They depend however on initial droplet size and composition.

Further on, Eqn. 27 from [16], describing evaporation of a droplet of a binary mixture in an atmosphere void of vapours (dry), can in the limit of infinitesimally low volatility of one of the components, also be applied to SDS/DEG solution. Thus, the temporal evolution of SDS/DEG solution droplet surface change rate can be described with a sigmoid (logistic) function.

In consequence, to deconvolute the influence of contaminating water we fitted the experimental data with a sigmoid function combined with the initial exponential decay. Then the exponential decay is subtracted from the experimentally measured surface change rates.

This procedure is illustrated in Fig. 5 for two droplets: (i) a droplet with the smallest initial SDS concentration – 20 mM and a small initial radius $a_0 = 9.9$ μm and (ii) a droplet with the biggest initial SDS concentration – 100 mM and the largest initial radius $a_0 = 16.54$ μm.

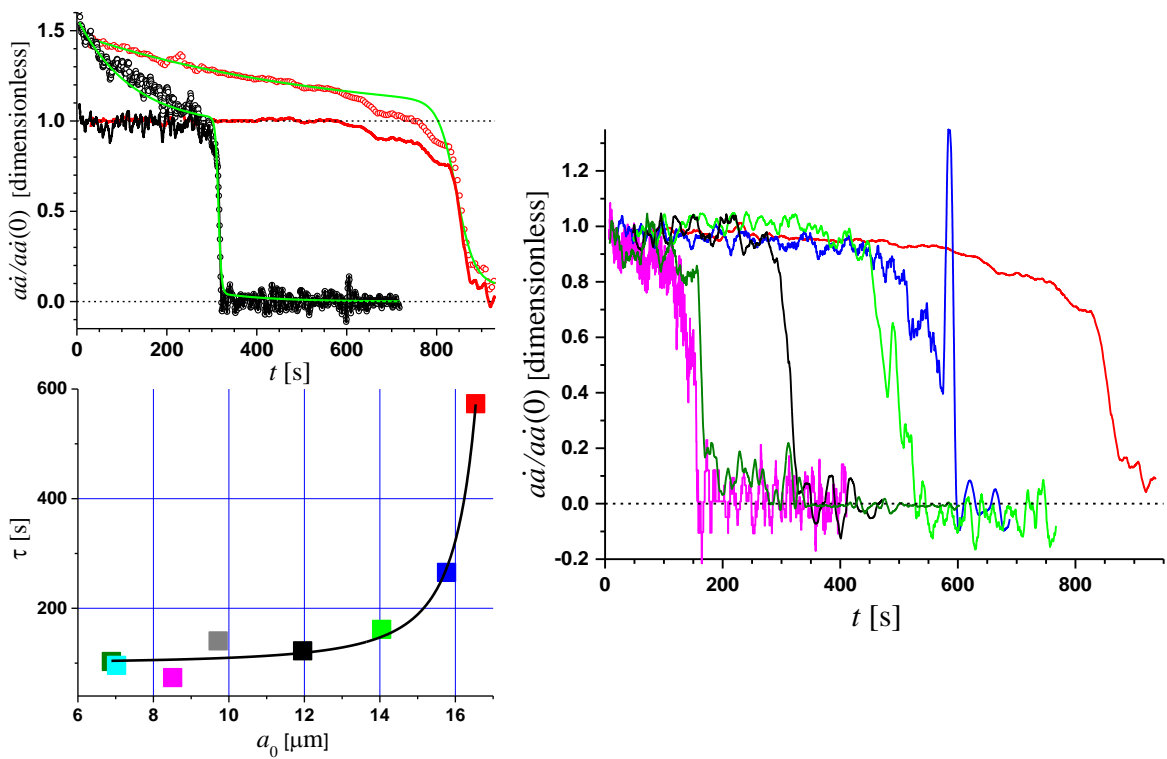

Fig.5. Top-left panel: examples of droplet surface change rate evolution – open black and red circles corresponding to droplets with initial SDS concentration of 20 and 100 mM respectively (most data points omitted for clearer visualization). Fitted sigmoid functions with the initial exponential decay are shown in green. The evolution of the surface change rate after subtraction of the influence of contaminating water and normalization is shown respectively in black and red solid lines. Bottom-left panel: the characteristic exponential decay times associated with water evaporation (water dwell times) *versus* the initial droplet radius for different initial SDS concentrations: red and olive – 100 mM, blue and magenta – 50 mM, green – 40 mM, black and grey – 20 mM. Black solid line – fitted exponential growth function. Right panel: examples of the evolution of droplet surface change rate, corresponding to Fig. 2, after subtraction of the influence of contaminating water and normalization.

It should be pointed out (Fig. 5 bottom-left panel) that water content dwell time $\tau$ visibly depends on the initial droplet radius, while its dependence on the initial SDS concentration is negligible. This is in agreement with the prediction of Eqn. 27 from [16] and the results shown in Fig. 5 from [16], and can be identified as the influence of non-stationary water-in-DEG diffusion (compare also [30], [31], [32]).

## 3.2. Evolution of surface layer

After deconvoluting the influence of contaminating water evaporation, the details of the evolution of SDS/DEG solution/suspension droplets can be accessed more conveniently.

In an atmosphere void of vapour (vapour density far from the droplet $\rho_\infty = 0$) the composite droplet evaporation rate is proportional to the density of vapour near (at) the droplet surface $\rho_a$ [33]:

$$a\dot{a}(t) = -\frac{D_{\text{DEG}}}{\rho_{\text{L}}}\rho_a \qquad (2)$$

where, in our case, the densities $\rho_\infty$ and $\rho_a$ pertain to DEG vapour and $D_{\text{DEG}}$ is the diffusion constant of DEG in air, while $\rho_{\text{L}}$ pertains to liquid (mixture) density. It should be noted here that in case of a SDS/DEG mixture, the density is fairly independent of (changing) composition: $\rho_{\text{L}} \cong \rho_{\text{SDS}} \cong \rho_{\text{DEG}} \cong const$.

The density of DEG vapour near the composite droplet surface must be expressed in terms of the saturated vapour density $\rho_{\text{sat}}$ and the composition of the droplet (surface). Since SDS/DEG mixture has a very complex polymorphic structure, exhibiting characteristics of both suspension and solution, we adapted the relation (6-33) from [33], being a modification of the Köhler equation

$$a\dot{a} = -D_{\text{DEG}}\frac{\rho_{\text{sat}}}{\rho_{\text{L}}}\exp\left(-K\frac{v_{\text{SDS}}}{1-v_{\text{SDS}}}\right), \qquad (3)$$

where $v_{\text{SDS}}$ is the SDS volume fraction. Thus, Eqn. 3 describes in an intuitive manner how the ratio of the SDS fraction $v_{\text{SDS}}$ and the DEG fraction ($v_{\text{DEG}} = 1 - v_{\text{SDS}}$) at/below the interface controls the DEG vapour pressure above the interface. Let's remark that in the case of SDS/DEG mixture volume fraction is equal to the mass fraction.

In case of a colloidal suspension consisting of a mixture of monomers and micelles such as SDS/DEG mixture, it is very difficult to predict the factor $K$ theoretically. In Appendix A, we propose a derivation of Eqn. 3 with a prediction of $K$ for somewhat idealized conditions given in Appendix B. On the other hand, $K$ is easy to find from the experimental data (Figs. 6 and B1).

Rather unexpectedly (see discussion in Appendix B), the experimental data for SDS/DEG mixture droplets could be reproduced – barring the modulations, which will be discussed in the following sections – with an evaporation rate calculated for $K = 1$ (Fig. 6). In order to study this result further, we have checked several different mixtures. However, our experimental data are still too sparse to assess how general these findings are (Fig. B1).

As it has been pointed out, the surface layer controls the evaporation rate and the SDS surface structure plays an important role. At this stage of analysis, we propose a general scenario, which we shall develop in the following sections.

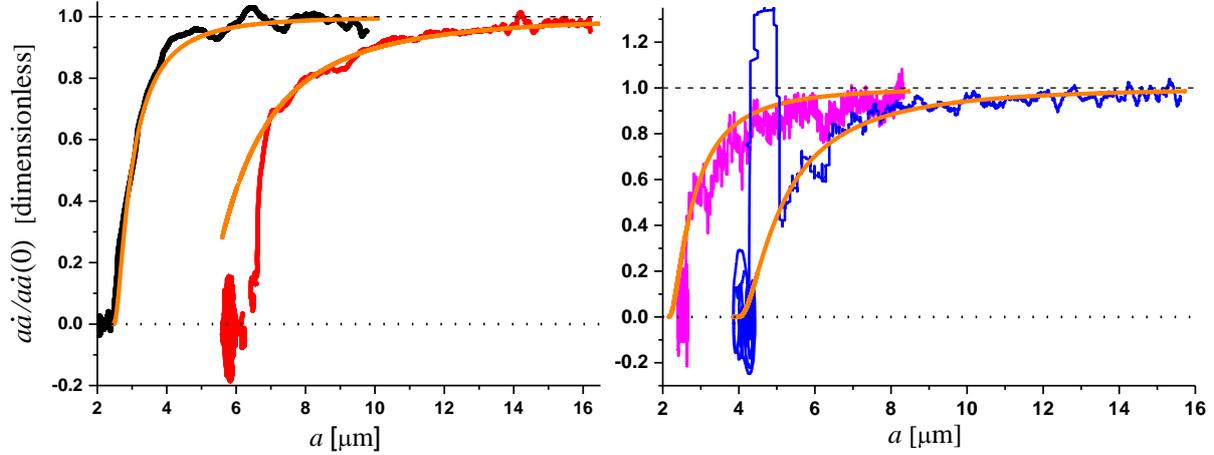

Fig. 6. The comparison of the calculated evaporation rate for $K = 1$ (solid orange lines) with the experimental results for SDS/DEG droplets. Left panel: the smallest droplet ($a_0 = 9.8$ μm) with the smallest initial SDS concentration (20 mM – black) and the biggest droplet ($a_0 = 16.54$ μm) with the highest initial SDS concentration (100 mM – red). Right panel: evaporation rates of a small ($a_0 = 8.53$ μm – magenta) and a big ($a_0 = 15.77$ μm – blue) droplet with initial SDS concentration of 50 mM. The rates were normalized to their initial values (at $t = 0$).

Using Eqn. 3 it is possible to determine the SDS volume fraction at the surface from the droplet radius evolution:

$$v_{SDS} = -\frac{\ln\frac{a\dot{a}(t)}{a\dot{a}(t=0)}}{K - \ln\frac{a\dot{a}(t)}{a\dot{a}(t=0)}} \quad (4)$$

In a system at equilibrium, for SDS fraction higher than the critical micelles concentration (CMC), the surface is considered to be covered with a monolayer of monomers with saturated concentration. However, as mentioned above, the evaporating composite droplet is not in equilibrium and various dynamic phenomena take place in the surface layer causing variations of $v_{SDS}$. This we investigate in the next sections.

### *3.3. Fine details of evaporation rate and surface fractions*

Looking closely at the details of evolution of the evaporation rate allows to follow the changes of the surface SDS fraction and helps to understand the behaviour of the surface of microdroplets.

The evolution of the surface SDS concentration generally follows the mean concentration in the volume (Fig. 7). However, its oscillatory character is very significant.

Oscillations are particularly well visible in the excess of the surface SDS volume fraction over its mean volume value $\Delta v_{SDS} = v_{SDS} - \overline{v_{SDS}}$, where

$$\overline{v_{SDS}} = \frac{V_{SDS}}{V_{SDS}+V_{DEG}}, \quad (5)$$

and $V_{SDS}$ and $V_{DEG}$ are the (total) volumes of SDS and DEG in the droplet (compare e.g. [34][35]).

The behaviour of the volume fraction excess $\Delta v_{SDS}$ can be compared to a pendulum, where the potential energy is transformed into the kinetic energy of motion and vice versa. Here the potential energy is associated with the compression (and warping) of the surface layer of SDS in the force field of the Laplace pressure. When the surface layer collapses, the energy is transformed into the kinetic energy of fluid flow (outflow and evaporation of DEG), resulting in consecutive compression and warping of the surface layer, and so on. The character of oscillations depends on the initial droplet radius and SDS concentration (volume fraction) (Fig. 7). To compare the fine details of evolution of droplets with different initial radii having the same (or very similar) initial density, we express the droplet radius by the SDS mean fraction $\overline{v_{SDS}}$. Since $\overline{v_{SDS}}$ changes by several orders of magnitude during the droplet drying, it is convenient to express it with an exponent $n$, where ($-n$) counts the instances of SDS $\overline{v_{SDS}}$ doubling:

$$\overline{v_{SDS}} = 2^{-n}, \text{ or } n = -\frac{\ln \overline{v_{SDS}}}{\ln 2}. \quad (6)$$

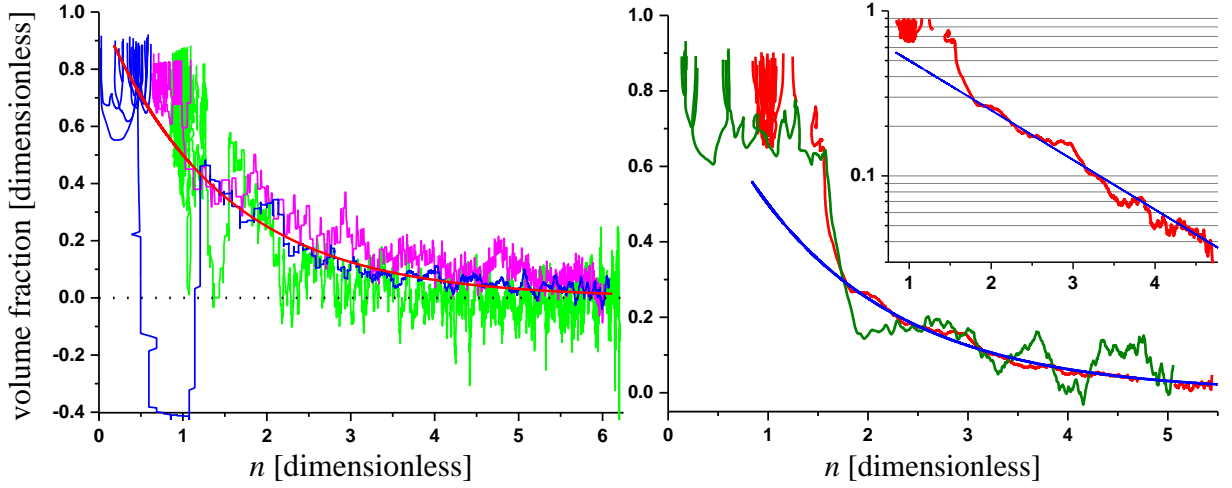

Fig. 7. Examples of SDS volume fraction evolution. Smooth curves (red and blue) – mean SDS volume fraction in the droplet volume $\overline{v_{SDS}} = 2^{-n}$, described with Eqn. 6. Other curves – surface $v_{SDS}$. Left panel – evolutions for different initial SDS concentration and initial radius: blue – 50 mM, 15.77 μm; green – 40 mM, 14.10 μm; magenta – 50 mM, 8.53 μm. Right panel: red – 100 mM, 16.54 μm; olive – 100 mM, 6.88μm. Inset: visualization of stepwise evolution in logarithmic scale.

Thus, increasing the natural number $n$ by 1 corresponds to halving SDS concentration, i.e. $n = 0$ for pure SDS, $n = 1$ for equal volumes of DEG and SDS $V_{DEG} = V_{SDS}$, etc.

The damping of oscillations seems to increase with the increase of the droplet initial radius. This can be observed for the droplets with comparable initial SDS mole fraction (40 and 50 mM), but different initial radii (Fig. 7, left panel).

Let us remark that the oscillations of $\Delta v_{SDS}$ (SDS excess) for smallest droplets (initial radius of 6.88 μm and 8.53 μm) presented in Fig. 8 have constant amplitude and $v_{SDS}$ does not change its sign.

For larger droplets the oscillations exhibit strong damping with parts of the evolution showing negative $\Delta v_{SDS}$. This can be interpreted as a growth of evaporating liquid (DEG) fraction in the surface layer. It seems that in this case the consecutive collapses of the surface layer are caused by fracturing of the "crystalized" SDS forms on the surface. This leads to acceleration of DEG evaporation due to its outflow via the capillary effect [36]. The extreme case of the phenomenon can be spotted as an "explosion" on the blue ($a_0 = 15.77$ μm, $c_{SDS} = 50$ mM) and the green ($a_0 = 14.10$ μm, $c_{SDS} = 40$ mM) curve.

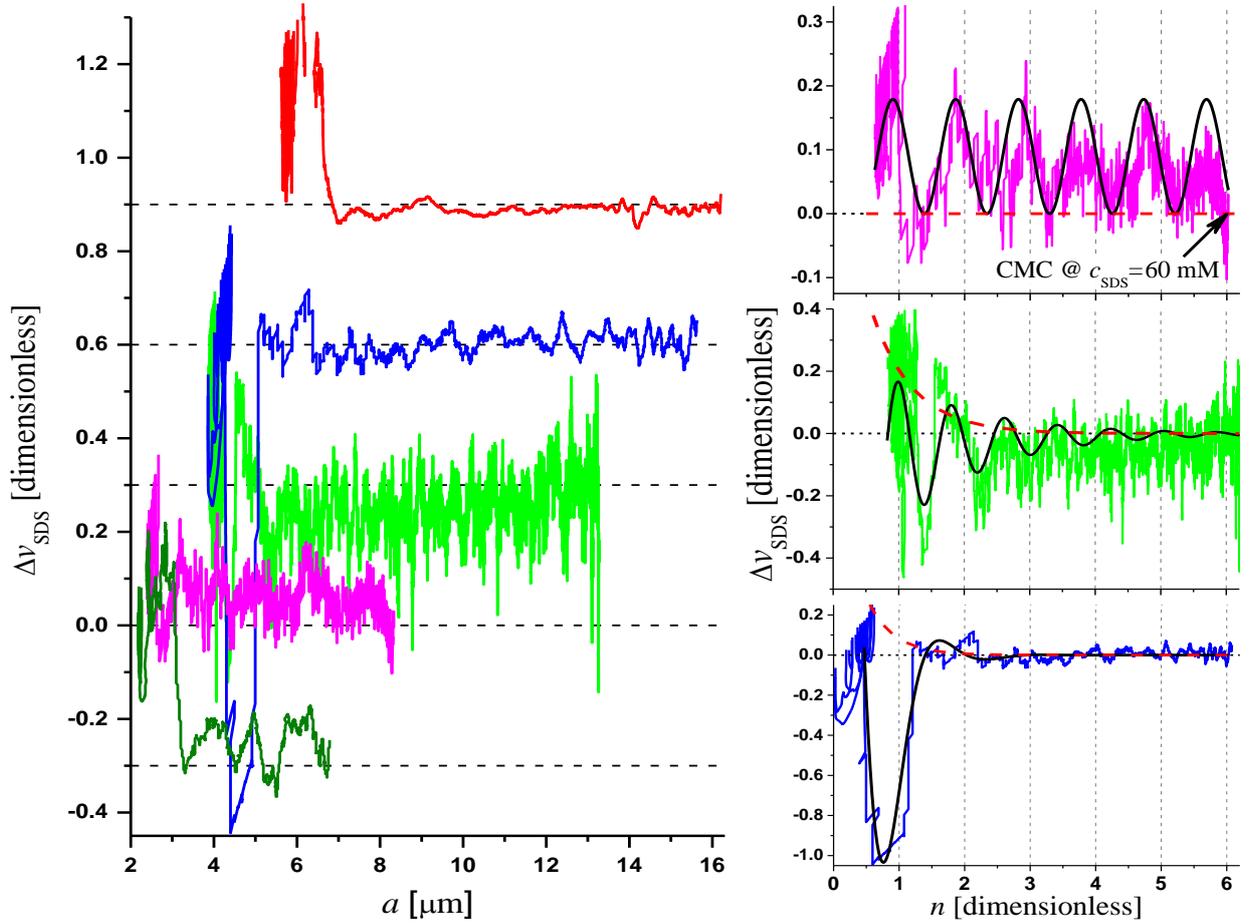

Fig. 8. Evolution of the SDS volume fraction excess $\Delta v_{SDS}$ as the function of droplet radius (left panel) and of $n$ – doubling of concentration (right panel). In the left figure, the zero levels were shifted by: −0.3 (olive; $a_0 = 6.88$ μm, $c_{SDS} = 100$ mM), 0.3 (green; $a_0 = 14.10$ μm, $c_{SDS} = 40$ mM), 0.6 (blue; $a_0 = 15.77$ μm, $c_{SDS} = 50$ mM) and 0.9 (red; $a_0 = 16.54$ μm, $c_{SDS} = 100$ mM) to increase visibility. The black solid lines in the right figure are fits visualizing the (damped) oscillations. The red dashed lines visualize the dumping of oscillations, modelled with the exponential decay $\exp(-n/\gamma)$, where $\gamma = 0$, $0.63\pm0.02$ and $0.4\pm0.007$ for magenta, green and blue curves respectively. The (damped) oscillation frequencies are $6.67\pm0.02$, $7.38\pm0.05$ and $4.04\pm0.02$ respectively.

"Crystallization" and fracturing of the surface layer manifests for larger droplets having rather higher curvature radius. This effect diminishes with the (initial) concentrations of SDS, so that for the droplets with initial concentration of 100 mM it is not visible. It seems to suggest that in that case "crystallization" occurs throughout the droplet volume, which prevents collapsing of the surface layer.

## 4. Multiple Critical Concentrations (MCC)

In order to explain oscillations of the surface SDS volume fraction, like those seen in Fig. 8, we propose the following scenario.

For SDS concentration smaller than CMC (practically outside the scope of our experiment) the surface film of SDS monomers behaves as a Langmuir monolayer compressed due to the droplet evaporation – shrinking of the droplet surface. At CMC, the surface monomer concentration saturates and further evolution of the surface film due to the droplet evaporation is possible only in radial direction. The surface film must buckle-in and form bulges evolving into vesicles, when the surface area continues to shrink [37], [38], [39]. In a way, the process of formation of vesicles at the surface by buckling can be perceived as a nucleation due to the supersaturation of the surface SDS concentration. The process of (heterogeneous) nucleation should be initiated by nuclei of condensation.

It can be imagined, and is hinted by the experimental results discussed below, that the existing vesicles (micelles) serve as these condensation nuclei. If each vesicle initiates the formation of another, the number of vesicles should double whenever a new generation of vesicles arises.

The bucklings start to transform into the vesicles at maxima of surface volume fraction excess. Then vesicles are submerged by the surface tension and finally they are dispersed in the droplet volume. The surface returns to equilibrium with the volume, which corresponds to the surface covered with saturated SDS monolayer as it is at CMC. Since the number of vesicles doubled due to the nucleation at the surface, the equilibrium between monomers and vesicles in the volume is reached for doubled monomer density, as approximatelly half of the SDS mass is in the form of monomers and half in the form of vesicles/micelles [40].

Therefore the consecutive point of equilibrium should occur (approximatly) for the doubled SDS fraction. This is true only if the mean vesicle/micelle composition (size) stays (approximately) constant for growing SDS concentration [41], [42].

Further droplet evaporation causes the repetition of the process of buckling and production of new generations of vesicles. Since not all the vesicles can participate in this process, a reproducibilty parameter α must be introduced. If all vesicles are serving as condensation nuclei, $\alpha = 1$.

After the $k$-th cycle, the SDS concentration at equlibrium $n(k)$ is $2^{\alpha(k-1)}$ times higher than the concentration at CMC:

$$\frac{n(k)}{n(\text{CMC})} = 2^{\alpha(k-1)}.$$

This can be expressed in terms of SDS volume fraction:

$$\frac{\overline{v_{\text{SDS}}}(k)}{\overline{v_{\text{SDS}}}(\text{CMC})} = 2^{\alpha(k-1)} \tag{7}$$

The experimental values of points of equilibrium $\overline{v_{\text{SDS}}}(k)$, corresponding to the right-top panel of Fig. 8, are shown in the left panel of Fig. 9. Although the surface-volume SDS concentration equilibria can be slightly shifted in respect to natural-numbered values of $n$, still fitting of these points enables determination of the critical micelles concentration. Or to be more precise – a manifold of critical micelles concentrations. The fit of Eqn. 7, shown in the left panel of Fig. 9, yields $\overline{v_{\text{SDS}}}(\text{CMC}) = 0.0172 \pm 0.0021$ and the efficiency of vesicles reproduction is $\alpha = 0.83 \pm 0.05$. Therefore, the CMC for SDS in DEG is 60±2 mM.

Obviously, the dependence $\overline{v_{\text{SDS}}} = 2^{-n}$ has a straightforward relation to $k$ – the number of the cycle of surface-volume SDS concentration equilibrium, or zero of surface SDS fraction excess $\Delta v_{\text{SDS}}$. The relation between the parameters $k$ and $n$ (doubling/halving of SDS concentration) is shown in the right panel of Fig. 9.

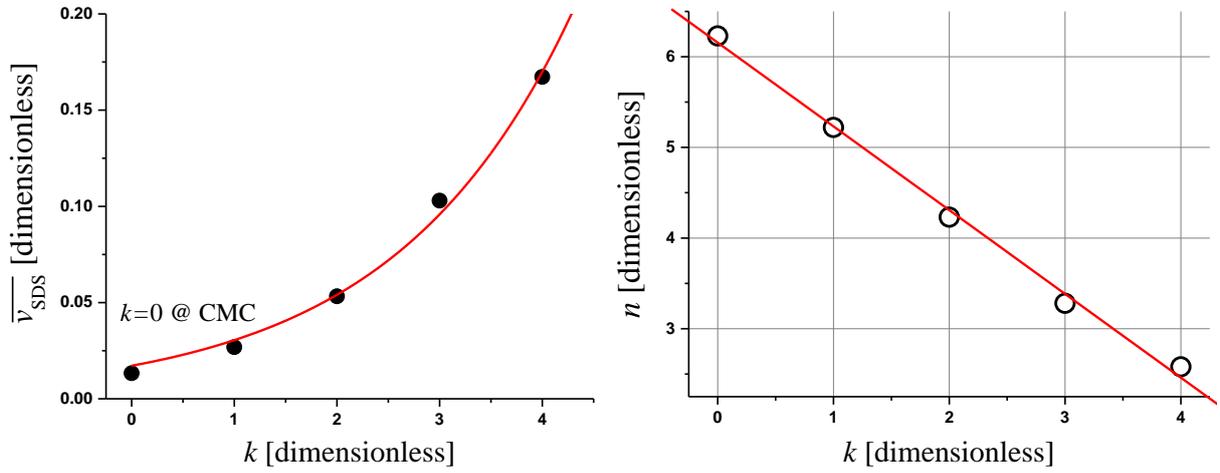

Fig. 9. Left panel: black circles – the experimental values of points of surface-volume SDS concentration equilibrium $\overline{v_{\text{SDS}}}(k)$, corresponding to the right-top panel of Fig. 8; red solid line – fit of Eq. (7). Right panel: the relation between the parameters $k$ and $n$ (doubling/halving of SDS concentration). Linear fit shown in red line.

It should be stressed, that at these critical points consecutive generations of new micelles are created at the surface. The other critical concentrations (CC), encountered for larger droplets, are related to points of the collapse of ("crystalised") surface layers, i.e. maxima of SDS surface concentration. These CC can be understood as defining surface pressure needed to elicit layer collapse.

## 5. Conclusions

In this paper, we studied evaporation of single microdroplets of mixtures of SDS /DEG.

First, we investigated the evaporation rate of microdroplets, proportional to the rate of change of the surface area. Observed evaporation rates did not follow so-called "a-square law" neither for very small nor for high, approaching dense and dry SDS, initial SDS concentration.

The investigation of small density region made it possible to conclude that the observed evolution of rates is influenced by the evaporation of residual water. The process was approximately described with exponential decay of evaporation rate, with experimentally determined decay rate ($1/\tau$) being dependent on droplet radius rather than density.

Recognition of the influence of water enables further study of the evolution of droplet.

At this stage we obtained an interesting approximation, making possible to extract the surface volume fraction of SDS and its evolution from the evaporation rate. Our results indicate that the evaporation rate of DEG/SDS mixture is controlled just by the exponent of the ratio of fractions of the components at the surface $\exp\left(-K\frac{v_{\text{SDS}}}{v_{\text{DEG}}}\right)$. It seems important to stress that the coefficient $K$ was found equal to 1 not only for DEG/SDS mixtures but also for several other colloidal suspensions and mixtures that we studied.

The knowledge of the coefficient $K$ and the influence of water enabled the study of the surface SDS density.

In our experimental results, we identified two modes of the SDS surface density evolution dependent on the initial droplet radius and composition (best seen in Fig. 7). First, with the growth of the initial droplet radius, a fairly smooth evolution exhibiting slow oscillations transforms into evolution with quick and complex oscillations. For the largest droplet (15 μm) and the highest initial density, the droplet evolution exhibits only slight oscillations.

We suggest that the slow oscillations of the fairly smooth evolution are connected with the surface SDS (mono)layer compression, leading to the successive monolayer collapses (nucleation of vesicles) and doubling of the number of vesicles in the droplet. Such an interpretation made it possible to determine the multiple CMC densities and the CMC value. As a by-product, we found the vesicle reproduction coefficient $\alpha$.

The evolution of larger droplets, with its rapid changes of surface density, suggests the presence of crystalline forms at the surface, which enables an increase of the evaporation rate when these forms are fractured. A rapid increase of the evaporation ensues then due to the capillary phenomena arising in the fractures of the surface layer.

**Appendix A**

In case of a droplet of a mixture, $\rho_a$ must account for the droplet composition. When the droplet is large enough (> ~100 nm) and the mixture is ideal – as we assumed in [16] – the vapour density of $i$-th component over the droplet surface is described by Raoult's law

$$\rho_{a,i} = \rho_{\text{sat},i} x_i, \tag{A.1}$$

where $\rho_{\text{sat},i}$ and $x_i$ are the $i$th component saturated vapour density and mole fraction respectively. However, when we allow that mixture is non-ideal, we must introduce an activity coefficient $\gamma_i$ into the Raoult's law:

$$\rho_{a,i} = \rho_{\text{sat},i} x_i \gamma_i. \tag{A.2}$$

Then the Gibbs-Duhem equation can be applied. In case of SDS/DEG mixture it takes the form:

$$N_{SDS}d(\ln(x_{SDS}\gamma_{SDS})) = -N_{DEG}d(\ln(x_{DEG}\gamma_{DEG})) \tag{A.3}$$

We assume that $\gamma_{SDS} \cong const$ and notice that for the most part of the droplet evolution $N_{SDS}/N_{DEG} \cong x_{SDS}$. Obviously, when $x_{SDS} \to 0$, $x_{DEG}\gamma_{DEG} = 1$. Then, after integration of (A.3) we obtain:

$$x_{DEG}\gamma_{DEG} = \exp(-x_{SDS}) \tag{A.4}$$

and in consequence

$$a\dot{a} = -D_{DEG}\frac{\rho_{sat}}{\rho_L}\exp(-x_{SDS}) \tag{A.5}$$

It must be underlined that $x_{SDS}$ corresponds to the (local) mole fraction in the region where evaporation takes place, i.e. at/near the droplet surface.

## Appendix B – *Approximate SDS radial distribution*

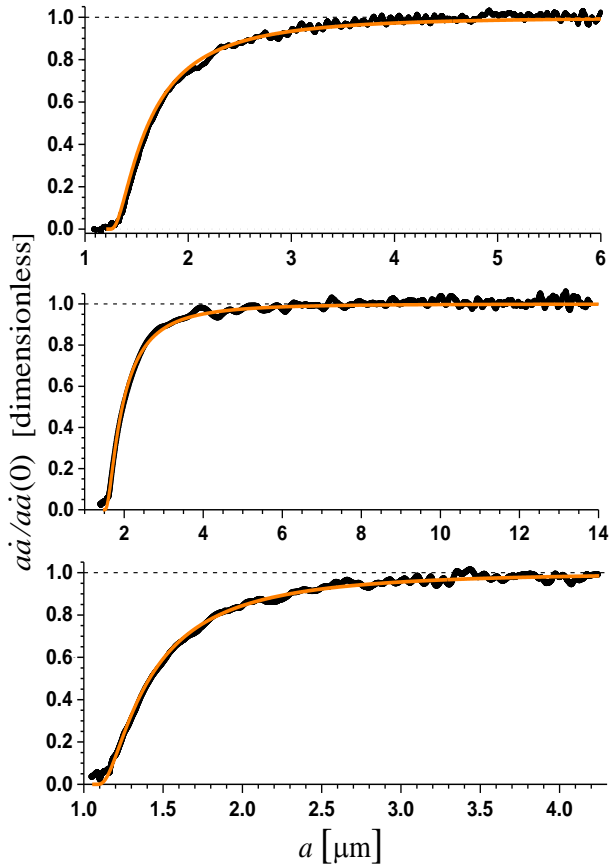

Fig. B1. The comparison of the calculated evaporation rate for $K = 1$ (solid orange line) with the experimental results for suspensions of different nanospheres in DEG. Top panel: Ag, 10 nm diameter; middle panel: Au, 250 nm diameter; bottom panel: SiO$_2$, 450 nm diameter.

As it has been mentioned above, during the evaporation the droplet is not in equilibrium and, as long as there are no mechanisms of mixing between the volume and the surface, e.g. due to a capillary flow [43], [44] or convection [45], there is a radial distribution of SDS in the droplet and the concentration of SDS at the surface can be significantly higher than its mean concentration. It seems reasonable to assume [46], [16]] that the radial profile of the distribution is constant, while its local value changes along with the average SDS concentration in the droplet: $x_{SDS} \propto \overline{x_{SDS}}$.

Then, the surface value of the mole fraction can be expressed as

$$x_{SDS} = \widetilde{K}\frac{V_m^{DEG}}{V_m^{SDS}}\frac{a_\infty^3}{a^3 - a_\infty^3}, \tag{B.1}$$

where $a_\infty$ is the radius of a final dry object formed from the SDS/DEG droplet, $V_m^i$ is the molar volume of the *i*-th component and $\widetilde{K}$ is a constant.

It can be rightly assumed that the increase of SDS concentration near the surface –

both SDS monomers and micelles – is driven mostly by the moving surface of an evaporating droplet. The effect of this driving force can be visualised as the radial pressure gradient causing a certain buoyancy pushing objects with higher molar volume (rather than density, e.g. micelles, due to the non-gravitational nature of the force) towards the centre of the droplet. It might be expected that due to the effect of this force the total volumes within the near-surface layers, occupied by components with different molar volumes, will be equal:

$$\delta N_{\text{DEG}} V_m^{\text{DEG}} = \delta N_{\text{SDS}} V_m^{\text{SDS}}, \tag{B.2}$$

where $\delta N_i$ is the number of molecules (moles) of a component $i$ in the layer. This means that

$$x_{\text{SDS}} \propto \frac{V_m^{\text{DEG}}}{V_m^{\text{SDS}}}, \tag{B.3}$$

which suggests that indeed we can assume (compare Figs. 6 and B1)

$$\widetilde{K} \frac{V_m^{\text{DEG}}}{V_m^{\text{SDS}}} \equiv K = 1 \tag{B.4}$$

**References**

[1] Rieger M. Surfactants in Cosmetics. CRC Press; 2017.

[2] Norn V. Emulsifiers in Food Technology. Wiley; 2015.

[3] Porter MR. Handbook of Surfactants. Springer US; 2013.

[4] Robb ID. Specialist Surfactants. Springer Netherlands; 2012.

[5] Shah DO. Micelles: Microemulsions, and Monolayers: Science and Technology. CRC Press; 2018.

[6] Siebert TA, Rugonyi S. Influence of liquid-layer thickness on pulmonary surfactant spreading and collapse. Biophys J 2008;95:4549–59. https://doi.org/10.1529/biophysj.107.127654.

[7] Kawai T, Kamio H, Kondo T, Kon-No K. Effects of concentration and temperature on SDS monolayers at the air - Solution interface studied by infrared external reflection spectroscopy. J Phys Chem B 2005;109:4497–500. https://doi.org/10.1021/jp046858i.

[8] Burlatsky S, Atrazhev V, Dmitriev D, Sultanov V, Timokhina E, Ugolkova E, et al. Surface tension model for surfactant solutions at the critical micelle concentration. J Colloid Interface Sci 2013;393:151–60. https://doi.org/10.1016/j.jcis.2012.10.020.

[9] Kralchevsky PA, Danov KD, Anachkov SE, Georgieva GS, Ananthapadmanabhan KP. Extension of the ladder model of self-assembly from cylindrical to disclike surfactant micelles. Curr Opin Colloid Interface Sci 2013;18:524–31. https://doi.org/10.1016/j.cocis.2013.11.002.

[10] Fendler JH, Fendler EJ. Preface. In: Fendler JH, Fendler EJ, editors. Catal. Micellar

Macromoleular Syst., Academic Press; 1975, p. xi–xii. https://doi.org/https://doi.org/10.1016/B978-0-12-252850-7.50004-6.

[11] Malek SMA, Poole PH, Saika-Voivod I. Thermodynamic and structural anomalies of water nanodroplets. Nat Commun 2018;9:2402. https://doi.org/10.1038/s41467-018-04816-2.

[12] Ybert C, Lu W, Möller G, Knobler CM. Collapse of a monolayer by three mechanisms. J Phys Chem B 2002;106:2004–8. https://doi.org/10.1021/jp013173z.

[13] Lee KYC. Collapse Mechanisms of Langmuir Monolayers. Annu Rev Phys Chem 2008;59:771–91. https://doi.org/10.1146/annurev.physchem.58.032806.104619.

[14] Moroi Y, Rusdi M, Kubo I. Difference in surface properties between insoluble monolayer and adsorbed film from kinetics of water evaporation and BAM image. J Phys Chem B 2004;108:6351–8. https://doi.org/10.1021/jp0306287.

[15] Davies JF, Haddrell AE, Reid JP. Time-resolved measurements of the evaporation of volatile components from single aerosol droplets. Aerosol Sci Technol 2012;46:666–77. https://doi.org/10.1080/02786826.2011.652750.

[16] Kolwas M, Jakubczyk D, Do Duc T, Archer J. Evaporation of a free microdroplet of a binary mixture of liquids with different volatilities. Soft Matter 2019;15:1825–32. https://doi.org/10.1039/C8SM02220H.

[17] Wu X, Wu Y, Saengkaew S, Meunier-Guttin-Cluzel S, Gréhan G, Chen L, et al. Concentration and composition measurement of sprays with a global rainbow technique. Meas Sci Technol 2012;23:1–13. https://doi.org/10.1088/0957-0233/23/12/125302.

[18] Wu Y, Crua C, Li H, Saengkaew S, Mädler L, Wu X, et al. Journal of Quantitative Spectroscopy & Radiative Transfer Simultaneous measurement of monocomponent droplet temperature / refractive index , size and evaporation rate with phase rainbow refractometry. J Quant Spectrosc Radiat Transf 2018;214:146–57. https://doi.org/10.1016/j.jqsrt.2018.04.034.

[19] Wilms J, Weigand B. Composition measurements of binary mixture droplets by rainbow refractometry. Appl Opt 2007;46:2109–18.

[20] Onofri FRA, Ren KF, Sentis M, Gaubert Q, Pelcé C. Experimental validation of the vectorial complex ray model on the inter-caustics scattering of oblate droplets. Opt Express 2015;23:15768. https://doi.org/10.1364/oe.23.015768.

[21] Cabral JT, Adamo M, Poulos AS, Lopez CG. Droplet microfluidic SANS. Soft Matter 2018;14:1759. https://doi.org/10.1039/c7sm02433a.

[22] Jakubczyk D, Derkachov G, Kolwas M, Kolwas K. Combining weighting and scatterometry: Application to a levitated droplet of suspension. J Quant Spectrosc Radiat Transf 2013;126:99–104. https://doi.org/10.1016/j.jqsrt.2012.11.010.

[23] Osswald TA, Baur E, Brinkmann S, Oberbach K, Schmachtenberg E, Osswald TA, et al. International Plastics Handbook. Int. Plast. Handb., 2006. https://doi.org/10.3139/9783446407923.fm.

[24] Lunkenheimer K, Czichocki G. On the Stability of Aqueous Sodium Dodecyl Sulfate Solutions. J Colloid Interface Sci 1993;160:509–10. https://doi.org/10.1006/jcis.1993.1429.

[25] Woźniak M, Archer J, Wojciechowski T, Derkachov G, Jakubczyk T, Kolwas K, et al. Application of a linear electrodynamic quadrupole trap for production of nanoparticle aggregates from drying microdroplets of colloidal suspension. J Instrum 2019;14:P12007--P12007. https://doi.org/10.1088/1748-0221/14/12/p12007.

[26] Kolwas M, Kolwas K, Derkachov G, Jakubczyk D. Surface diagnostics of evaporating droplets of nanosphere suspension: Fano interference and surface pressure. Phys Chem Chem Phys 2015;17. https://doi.org/10.1039/c5cp00013k.

[27] Videen G, Bickel WS. Light-scattering resonances in small spheres. Phys Rev A 1992;45:6008–12. https://doi.org/10.1103/PhysRevA.45.6008.

[28] Hergert W, Wriedt T. The Mie Theory. vol. 169. Berlin, Heidelberg: Springer Berlin Heidelberg; 2012. https://doi.org/10.1007/978-3-642-28738-1.

[29] Jakubczyk D, Kolwas M, Derkachov G, Kolwas K, Zientara M. Evaporation of microdroplets: The "radius-square-law" revisited. Acta Phys Pol A 2012;122. https://doi.org/10.12693/APhysPolA.122.709.

[30] Whitaker S. Simultaneous Heat, Mass, and Momentum Transfer in Porous Media: A Theory of Drying. In: Hartnett JP, Irvine TFBT, editors. Adv. Heat Transf., vol. 13, Elsevier; 1977, p. 119–203. https://doi.org/10.1016/S0065-2717(08)70223-5.

[31] Handscomb CS, Kraft M, Bayly AE. A new model for the drying of droplets containing suspended solids. Chem Eng Sci 2009;64:628–37. https://doi.org/10.1016/j.ces.2008.04.051.

[32] Handscomb CS, Kraft M, Bayly AE. A new model for the drying of droplets containing suspended solids after shell formation. Chem Eng Sci 2009;64:228–46. https://doi.org/10.1016/j.ces.2008.10.019.

[33] Pruppacher HR, Klett JD. Microphysics of Clouds and Precipitation. vol. 18. Dordrecht: Springer Netherlands; 2010. https://doi.org/10.1007/978-0-306-48100-0.

[34] Mitropoulos AC. What is a surface excess? J Eng Sci Technol Rev 2008;1:1–3. https://doi.org/10.25103/jestr.011.01.

[35] Adamson AW, Adamson TA, Gast AP. Physical Chemistry of Surfaces. Wiley; 1997.

[36] Dong L, Johnson D. Surface Tension of Charge-Stabilized Colloidal Suspensions at the Water−Air Interface. Langmuir 2003;19:10205–9. https://doi.org/10.1021/la035128j.

[37] Shi W. The structure and dynamics of Nano Particles encapsulated by the SDS monolayer collapse at the water/TCE interface. Sci Rep 2016;6:37386. https://doi.org/10.1038/srep37386.

[38] Phan MD, Lee J, Shin K. Collapsed States of Langmuir Monolayers. J Oleo Sci 2016;65:385–97. https://doi.org/10.5650/jos.ess15261.

[39] Baoukina S, Monticelli L, Risselada HJ, Marrink SJ, Tieleman DP. The molecular

mechanism of lipid monolayer collapse. Proc Natl Acad Sci U S A 2008;105:10803–8. https://doi.org/10.1073/pnas.0711563105.

[40] Nagarajan R, Ruckenstein E. Theory of Surfactant Self-Assembly: A Predictive Molecular Thermodynamic Approach. Langmuir 1991;7:2934–69. https://doi.org/10.1021/la00060a012.

[41] Pisárčik M, Devínsky F, Pupák M. Determination of micelle aggregation numbers of alkyltrimethylammonium bromide and sodium dodecyl sulfate surfactants using time-resolved fluorescence quenching. Open Chem 2015;13. https://doi.org/10.1515/chem-2015-0103.

[42] Duplâtre G, Ferreira Marques MF, Da Graça Miguel M. Size of sodium dodecyl sulfate micelles in aqueous solutions as studied by positron annihilation lifetime spectroscopy. J Phys Chem 1996;100:16608–12. https://doi.org/10.1021/jp960644m.

[43] Bear J. Dynamics of fluids in porous media. New York (N.Y.) : Dover; 1988.

[44] Handscomb CS, Kraft M, Bayly AE. A new model for the drying of droplets containing suspended solids. Chem Eng Sci 2009;64:628–37. https://doi.org/https://doi.org/10.1016/j.ces.2008.04.051.

[45] Zhang N. Surface tension-driven convection flow in evaporating liquid layers. Surf. Tens. Flows Appl., vol. 661, 2006, p. 147–170.

[46] Dombrovsky LA, Sazhin SS. A Parabolic Temperature Profile Model for Heating of Droplets. J Heat Transfer 2003;125:535–7. https://doi.org/10.1115/1.1571083.